\documentclass[aps,reprint]{revtex4-1}
\usepackage{graphicx,color}
\usepackage{amssymb}   
\usepackage{amsmath}
\usepackage{upgreek}
\usepackage{hyperref}
\usepackage{bm}
\usepackage[normalem]{ulem}
\usepackage{float}

\begin{document}

\title{Spin supercurrent in two-dimensional superconductors 
with Rashba spin-orbit interaction}

\author{James Jun He$ ^{1,*, \dagger} $, Kanta Hiroki$ ^{2,\dagger} $, Keita Hamamoto$ ^{2} $, Naoto Nagaosa$ ^{1,2} $}
	
\affiliation{$^1 $RIKEN Center for Emergent Matter Science (CEMS), Wako, Saitama 351-0198, Japan\\	
	$^2 $Department of Applied Physics, The University of Tokyo, Tokyo 113-8656, Japan\\
    $ ^*$Corresponding author.  E-mail address: jun.he@riken.jp\\
    $ ^\dagger $These authors contribute equally. }

%\date{\today}
	
\begin{abstract}
	
	Spin current is a central theme in spintronics, and its generation is a keen issue. The spin-polarized current injection from the ferromagnet, spin battery, and spin Hall effect have been used to generate spin current, but Ohmic currents in the normal state are involved in all of these methods. On the other hand, the spin and spin current manipulation by the supercurrent in superconductors is a promising route for dissipationless spintronics. Here we show theoretically that, in two-dimensional superconductors with Rashba spin-orbit interaction, the generation of dissipationless bulk spin current by charge supercurrent becomes highly efficient, exceeding that in normal states in the dilute limit, i.e. when the chemical potential is close to the band edge, although the spin density becomes small there. This result manifests the possibility of creating new spintronic devices with long-range coherence.

\end{abstract}

%\pacs{}
\maketitle

%\section{Introduction}	
In spintronics, spin current plays an essential role to transfer the information associated with the spin degrees of freedom. Therefore the generation of spin current is an important issue, and several methods have been proposed and experimentally verified \cite{Maekawa}. 	
Various methods, such as the spin polarized current injection from the ferromagnet \cite{Datta,Wees,Weees2}, spin battery \cite{Saitoh,Ando,Dushenko, Lesne, Kondou}, and spin Hall effect \cite{Hirsch,Zhang,Murakami,Sinova,Kato,Wunderlich,Valenzuela}, have been employed to generate the spin current.
It has been proposed also that the spin current second order 
in the electric field can be generated in 
noncentrosymmetric systems with 
spin-orbit interaction \cite{Yu,Hamamoto}.
The spin currents mentioned above are either carried by itinerant electrons through dissipative electric currents, or by localized magnetic moments through exchange interactions. However, there can also be non-dissipative spin current in itinerant-electron systems. In this case, it is an equilibrium spin current without dissipation \cite{Rashba2003}. Although such a spin current is detectable according to Sonin \cite{Sonin}, it does not contribute to the transport property in a set up where the spin current can flow in and out. 	

On the other hand, superconducting spintronics is an emerging field attracting recent interest \cite{Grein,Alidoust,Bergeret,Timm,Linder4,JacobNP,Linder5,JacobPRB,Silaev1,Silaev2,Tokatly2014,Alidoust2016,Tokatly2017,Babaev}.  Although the spin degrees of freedom are usually quenched in singlet superconductors, the triplet component can be finite in noncentrosymmetric superconductors, ferromagnet-superconductor hybrids and Josephson junctions, or odd-parity superconductors, where the spins become (partially) active.   Therefore, it is an intriguing issue if one can generate a spin current in superconductors with zero or small dissipation. Actually, the spin supercurrent has been discussed in He3 related to the internal degrees of freedom \cite{Leggett}. 
Recently, Leurs et al. \cite{Zaanen} have reexamined the spin supercurrent in spin-orbit coupled systems, and classified it to the coherent and noncoherent parts, only the latter of which contributes to the continuity equation of the spin density and its generation or manipulation is the focus of superconducting spintronics. 

The superconducting magnetoelectric effect \cite{Edelstein, Edelstein2005,Yip, Wenyu2019} is also a result of the spins of Cooper pairs being active.  For example, the spin density  (or magnetization) induced by a supercurrent has been discussed by Edelstein \cite{Edelstein} in the case where the Fermi energy is large compared to the spin splitting by the Rashba interaction. Although this is justified in many situations, there are also systems where the spin-orbit splitting is comparable or even larger than the Fermi energy, e.g., the interface between LaAlO$_3$ and SrTiO$_3$ \cite{Triscone}, where the electron density can be tuned by gating. Therefore, it is desirable to cover the wide range of parameters, e.g., chemical potential,	the strength of Rashba interaction, and temperature.

Here, we study spin density and spin current produced by a superconducting current in a two-dimensional superconductor with Rashba spin-orbit interaction for wide parameter regions. 
We investigate in detail the properties of spin current in this system and find that the spin current generation efficiency is comparable to or even larger than that of normal state when normalized with the charge current density. 
The spin current we discuss here corresponds to the noncoherent part in the classification by Leurs et al. \cite{Zaanen} and hence contributes to the spin accumulation, unlike the equilibrium spin current in normal systems \cite{Rashba2003}.  
Furthermore, it is different from those in other superconducting spintronic setups involving ferromagnet-superconductor hybrids where the spin degrees of freedom of Cooper pairs become active due to the presence of ferromagnets. In contrast, what we obtain here is a bulk spin supercurrent induced by charge supercurrent without time-reversal symmetry breaking of the ground state. The spin degrees of freedom become active because of the spin-orbit coupling and the flow of a Cooper pair condensate. 
Similar to normal spin currents, such a spin supercurrent may be detected by connecting to a material that shows inverse spin Hall effect and converts the injected spin currents into a transverse voltage. 
The study of the bulk spin supercurrent carried by Cooper pairs is a significant step towards non-dissipative spintronics.

\section*{Results}
\subsection*{Model}
A two-dimensional superconductor (SC) with Rashba spin-orbit interaction (SOI) can be described by the following Hamiltonian,
\begin{align}
\hat{H}_0 = & \sum_{\mathbf k} [ (\frac{\hbar^2 k^2}{2m}-\mu) c_{s,{\mathbf k}}^\dagger c_{s,{\mathbf k}} + \alpha  ( \boldsymbol{\upsigma}_{ss'} \times \hbar {\mathbf k}) \cdot \hat{z} c_{s,{\mathbf k}}^\dagger c_{s',{\mathbf k}} \notag\\
& +  \Delta c_{\uparrow,{\mathbf k}}^\dagger c_{\downarrow,-{\mathbf k}}^\dagger + h.c.] ,
\end{align}
where $ \alpha $ is the Rashba SOI strength, $ \Delta $ is the SC order parameter and $ \boldsymbol{\upsigma} = \hat{ \mathbf x} \sigma^x + \hat{\mathbf y} \sigma^y + \hat{\mathbf z} \sigma^z $ is a spatial vector of Pauli matrices. The constants $ m $ and $ \hbar $ are the electron mass and the Planck constant respectively. Summation over implicit indexes is assumed. 

\begin{figure}
	\includegraphics[width=3.2in]{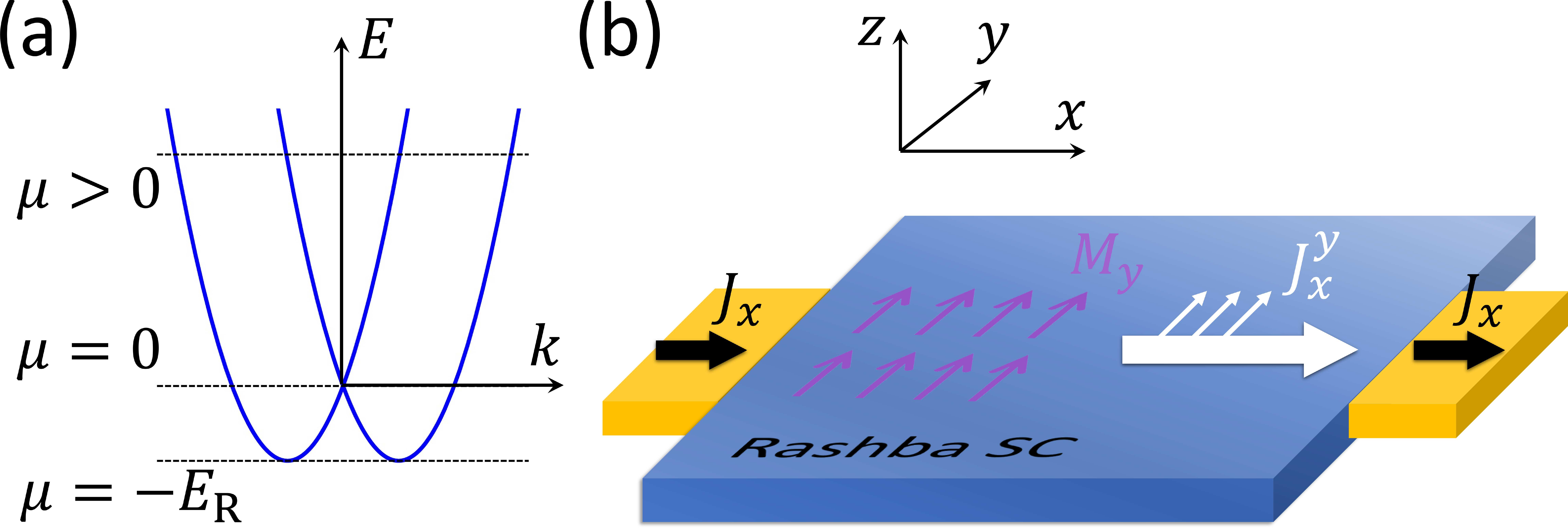}
	\caption{ {\bf Schematic pictures of the bands and the spin current.}  (a) The schematic normal state band structure of materials with Rashba splitting energy $ E_{\text R} $. $ E $ is the energy eigenvalue and $ k $ is the magnitude of the wave vector. The band edge is reached if the chemical potential $ \mu=-E_{\text R} $. (b) Cartoon picture showing the generation of magnetization $ M_y $ and spin current $ J_x^y $ by a charge supercurrent $ J_x $ in a two-dimensional superconductor (SC) with Rashba spin-orbit interaction. 
		%The spin current generated in the Rashba superconductor passes through a material with strong spin Hall effect, inverse process of which induces a transverse  voltage.
	}
	\label{FIG0}
\end{figure}

Without superconductivity, the normal-state band structure is schematically shown in FIG. \ref{FIG0}(a). When $ \mu \gg E_{\text R} $, $ E_{\text R}=m \alpha^2/2 $ being the Rasbha band splitting energy, the Fermi level is far above the band touching point. If $ \mu =0 $, the Fermi level cuts through this point and the inner Fermi surface shrinks to a point. The band bottom is reached when $ \mu = - E_{\text R} $. 

The superconductivity order parameter considered here is spin-singlet s-wave. In systems with Rashba spin-orbit interaction and superconductivity, s-wave may be mixed with p-wave and a general theory should include all of them. It has been shown that the actual amplitudes of the two different pairing potential depend on the form of the interaction \cite{Samokhin,Bauer}. When only even-parity interaction is considered, the p-wave order parameter vanishes. We assume pure s-wave order parameter in our calculation for simplicity.

Although we do not assume any spin-triplet order parameters in the Hamiltonian, the anomalous Green's function $ F_{\sigma \sigma'} (\omega,{\mathbf k})=(\psi_s+\mathbf d \cdot \boldsymbol{\upsigma} )i \sigma_y $, which carries the full information about the Cooper pairs, has both spin-singlet part $ \psi_s $ and spin-triplet components ${\mathbf d} $ when spin-orbit interaction is present \cite{Bauer}. As a result, the Cooper pairs become spin-active due to SOI even though the pairing potential is purely s-wave. When time reversal symmetry is broken, such as by the supercurrent $ J_x \sim q_x $ in our case, the spin of the Cooper pairs become polarized and the polarization is given by $ \mathbf S \sim i \mathbf d \times \mathbf d^* $ (more discussion in Supplementary Note 1). Since the pairs carry both polarized spin and momentum, they generate spin currents. This is illustrated schematically in FIG. \ref{FIG0}(b).

To investigate the spin and charge properties, it is convenient to introduce an SU(2) gauge field $\mathbf W^{\nu} \sigma^{\nu} $ \cite{Frohlich1,Frohlich2,Zaanen} along with the electromagnetic field $ \mathbf A $. Electrons coupled with both fields are described by the Hamiltonian
\begin{align}
\hat{H} =& \sum_{\mathbf k}   \{ [\frac{1}{2m}(\hbar\mathbf k + e \mathbf A - \frac{\hbar}{2} \mathbf W^\nu \sigma^\nu )^2-\mu ]_{ss'} c_{s,\mathbf k}^\dagger c_{s',\mathbf k} \notag\\
& +  \Delta c_{\uparrow,\mathbf k}^\dagger c_{\downarrow,-\mathbf k}^\dagger + h.c. \} .
\label{EqH}
\end{align}
Apparently, the Rashba term in $ \hat{H}_0 $ breaks SU(2) symmetry. It is actually equivalent to a constant gauge field (neglecting a numeric constant ) $\mathbf {\bar{ W}} =  (\bar{W}^x_y \sigma^x,\bar{W}^y_x \sigma^y,0)$, with $  \bar{W}^x_y =-\bar{W}^y_x= 2 m \alpha/\hbar $ \cite{Tokatly}. In this formulism, the spin current operator is conveniently defined as  $  \hat{\mathbf J}^\nu = \partial \hat{H}/\partial \mathbf W^\nu $. Or, starting from the free energy, the spin current can be obtained as $ \mathbf J^\nu = -\partial \mathcal{F}/\partial \mathbf W^\nu $.

\begin{figure}
	\includegraphics[width=3.2in]{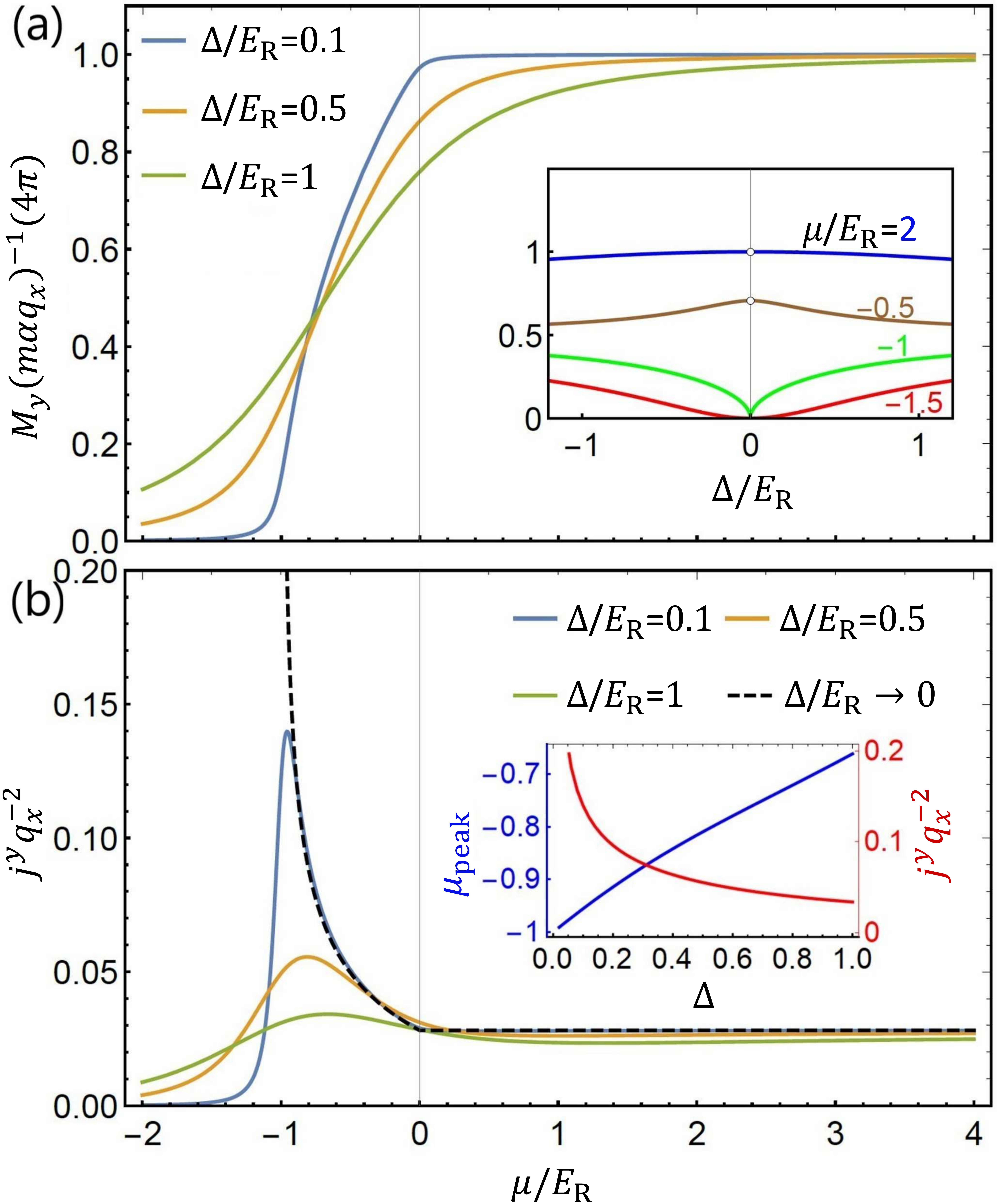}
	\caption{ {\bf Spin density and spin current density for a given Cooper pair wave vector $ q_x $.} (a) The spin density $ M_y $ as a function of the chemical potential $ \mu $.  The Rashba splitting energy $ E_{\text R} $ is used as the reference energy scale. (The explicit dependence on the spin-orbit interaction strength is shown in Supplementary Note 6.) Note that the band edge in the normal state is reached when $\mu=-E_{\text R}$ below which the normal state is insulating. The inset shows the dependence of the same quantity on the order parameter $ \Delta $ for chosen values of $ \mu $.  The quantity $ m $ is the electron mass and $\alpha  $ is the Rashba strength. (b)  The spin current as a function of chemical potential with various values of $ \Delta $. Peaks appear near the band edge.  The inset shows how the peak position $ \mu_{\text peak} $ and peak height $ j^y_{\text max}/q_x^2 $ change as we increase $ \Delta $. In all the results of this paper, we have set the electron charge $e=1$ and the Planck constant $\hbar=1$.}
	\label{FIG1} 
\end{figure}

When a uniform supercurrent is flowing (say along $ x $-direction), the SC order parameter acquires a constant phase gradient, i.e. $ \Delta = \Delta_0 \exp[2 i q_x x] $. Or, by a gauge transformation, it is equivalent to a constant vector potential $ A_x = \hbar q_x/ e $, which is effective only for SCs. 
Before showing the results, it is helpful to discuss the symmetry properties of the involved physical quantities. The spin current (spin density) is odd (even) under spatial inversion but even (odd) under time-reversal operation, while  charge current, or $ q_x $, is odd under both of them. Consequently, the lowest order of the spin current is $ J^\nu \sim q_x^2 $ and that of spin density is $ M_\nu \sim q_x $. Both of them must be odd functions of $ \alpha $.

\subsection*{ Spin density and spin current }
At zero temperature, the free energy $ \mathcal{F} $ is just the ground state energy. For arbitrary Rashba SOI strength and chemical potential, the spin polarization is along $ y $-direction and the density is (derivation in the Methods section)
\begin{align}
&M_y = \frac{ m \alpha q_x}{4\pi} g(\frac{\Delta}{E_{\text R}},\frac{\mu+E_{\text R}}{E_{\text R}}),\label{EqMy}\\
&g(d,u) = \frac{1}{2}- \frac{1}{2}\int_0^1 \frac{x^2-u}{\sqrt{d^2+\left(x^2-u\right)^2}} dx, \label{Eqg0}
\end{align}
where we have defined the Rashba splitting energy $ E_{\text R}=m \alpha^2/2 $. This is a generalized SC Edelstein effect \cite{Edelstein} in large-Rashba systems. The $ \mu $-dependence  of the spin density $ M_y $ is shown in FIG.\ref{FIG1}(a) with various values of the pairing potential. When $ \mu \gg E_{\text R} \gg \Delta$, we get $ M_y(\mu \gg E_{\text R}) = m \alpha q_x/4\pi $, in which the dependence on $\mu$ is absent. When $ \mu<0 $, the spin density gradually decreases as $\mu$ goes down.

In FIG.\ref{FIG1}(a), one readily notes that the spin density remains finite when $\Delta$ approaches zero. On the other hand, if $ \Delta=0 $, i.e. in a normal state, spin density due to the vector potential must vanish since one can trivially gauge out $ q_x $. Thus, there is a discontinuity in $M_y/q_x$ at $\Delta \rightarrow 0$. However, as we will see later, $ M_y $ actually vanishes when $ \Delta $ decreases to zero since the supercurrent $ J_x $ (and thus $ q_x $) also approaches zero, as discussed in Supplementary Note 2. Also, this discontinuity is absent at finite temperature, as shown in Supplementary Note 3. The $ \Delta $-dependence of the function $ g $ is shown in the inset of FIG. \ref{FIG1}(a), where a finite value is obtained at $ \Delta \rightarrow 0 $ for $ \mu/E_{\text R}>-1 $.  
In the limit $ \Delta/E_{\text R}\rightarrow 0 $, Eq.(\ref{EqMy}) simply becomes 
\begin{eqnarray}
M_y = \frac{ m \alpha q_x }{4\pi}\left\{ 
\begin{array}{ccc}
1, &\text{when $ \mu >0 $},\\
\sqrt{\mu/E_{\text R}+1}, &\text{when $ 0>\mu>-E_{\text R} $},\\
0 , &\text{when } \mu<-E_{\text R}.
\end{array}
\right.
\end{eqnarray}

Due to the form of the Rashba SOI, and as indicated by the  direction of the spin density obtained above, the only non-vanishing spin current induced by the supercurrent in $ x $-direction is  $ \mathbf J^y$. In the limit where  the chemical potential is high and the order parameter is small, i.e. $ \mu \gg E_{\text R} \gg \Delta $, the spin current density is  (derivation in the Methods section)
\begin{align}
\mathbf j^y_+ =  \alpha \hbar q_x^2 \hat{\mathbf x}/16\pi, \label{Eqjp}
\end{align}
and all other spin current components vanish. This expression is independent of $ \mu $ and $ \Delta $, similar to the spin density discussed previously. 

When $-E_{\text R}<\mu<0 $ and $ \Delta $ is small, the spin current density can be written as 
\begin{align}
\mathbf  j^y_- = \frac{\alpha \hbar q_x^2}{32\pi} f(\frac{\Delta}{E_{\text R}},\frac{\mu+E_{\text R}}{E_{\text R}}) \hat{\mathbf x},\label{Eqjm}
\end{align}
\begin{align}
f(d,u)  = -5 \sqrt{u} + \int_0^1 \frac{ d^2 (4 x^2+3)}{[d^2+(x^2-u)^2]^{3/2}} dx.
\end{align}
Particularly, when $ d=\Delta/E_{\text R} \rightarrow 0 $, the function $ f(d,u) $ simplifies to $ f(d \rightarrow 0,u) =  3/\sqrt{u} -\sqrt{u}$. Combined with Eqs.(\ref{Eqjm}), this leads to the same spin current density as Eq.(\ref{Eqjp}) when $ u =1 $ (corresponding to the band crossing point $ \mu=0 $). 
Also, the spin current	increases  because of the increase of the density of states when the chemical potential goes down until it reaches the band edge at $ \mu \rightarrow -E_{\text R} $ where it diverges, as shown by the dashed curve in FIG.\ref{FIG1}(b). 
 The two Fermi surfaces have the same (opposite) spin textures and the contributions from the two sub-bands add up (tend to cancel) when $ \mu $ is negative (positive).  
While the spin current is similar to spin density for large $ \mu $ in the sense that they both keep constant, it shows different behavior when $ \mu<0 $. 
Note that the $ \mu $-dependence of spin current resembles the density of states of the normal Rashba band, which also diverges at the band edge.

For general parameters, the spin current of the SC is calculated numerically and shown in FIG. \ref{FIG1}(b). Because of finite $ \Delta $, the aforementioned divergence of the spin current at the band edge is smoothed out while the spin current at $ \mu>0 $ is hardly changed even the pairing potential reaches  $ 0.5E_{\text R} $. The change of the  peak position  ($\mu_{\text peak}$) and peak height ($j^y_{\text max}/q_x^2$) as functions of $\Delta$  are shown in the inset of FIG.\ref{FIG1}(b). It turns out  that the peak position shifts almost linearly from the band edge when the order parameter increases while the height of the peak decreases in a nonlinearly.

\subsection*{ Bose-Einstein condensation regime} 
When the chemical potential is below the band edge, i.e. $\mu < -E_{\text R}$, the normal electronic state has no Fermi surface and SC cannot happen in the weak coupling limit of the Bardeen-Cooper-Schrieffer (BCS) theory. Assuming that the SC exists, it must be in  BEC regime  where electrons are tightly bond and the critical temperature is determined by the Bose-Einstein condensation (BEC) temperature \cite{Nozieres,QChen}. 
As shown in FIG.\ref{FIG1}, spin density and spin current quickly drops to zero when $\mu<-E_{\text R}$ if $\Delta$ is small. However, when the pairing potential is large and a BEC superconductor is achieved, they become finite.  Especially, when $ -u \gg d$ and  $ -u \gg 1 $, 
 a direct expansion of Eq.(\ref{Eqjm}) respect to $ d $ and $ u^{-1} $  gives
 $ g(d,u) \approx  \frac{1}{8} d^2/u^2=\frac{1}{8}\Delta^2 \mu^{-2}$ and $ \mathbf j^y_{BEC} = \alpha \hbar q_x^2  d^2 (12\pi)^{-1} | u|^{-3} \hat{\mathbf x}$. Both the spin density and spin current show power-law decay as $ \mu$ decreases. 
Note that the discontinuity at $ \Delta \rightarrow 0 $ does not appear when $ \mu<-E_{\text R} $.  

\subsection*{ Efficiency}
To relate our calculation to experiments, it is useful to convert the wave vector $q_x$ to the supercurrent density $j_x$. To calculate $ j_x $, we note that (charge) current  operator in a Rashba system is 
$
\hat{J}_x = e \int{d^2\mathbf k}[\frac{\hbar}{m} (k_x + q_x )- \alpha \sigma^y ]_{ss'} c_{s,\mathbf k}^\dagger c_{s',\mathbf k}.
$
At $ T=0 $, the usual paramagnetic term vanishes while the diamagnetic term remains. The last term proportional to $ \alpha $ contributes a supercurrent of $ e \alpha  M_y/\hbar $. So the zero temperature supercurrent density can be written as
\begin{align}
j_x
& = e \hbar q_x n_e(u) /m - 2 e \alpha  M_y/\hbar    
\notag\\
&=  [\hbar n_e(u) /m - g(d,u) E_{\text R} /(\pi\hbar)  ]  e q_x. \label{EqLondon}
\end{align}
The quantity $ n_e $ is the electron density. Eq.(\ref{EqLondon}) may be regarded as the generalized London equation for Rashba systems.

\begin{figure}
	\includegraphics[width=3.2in]{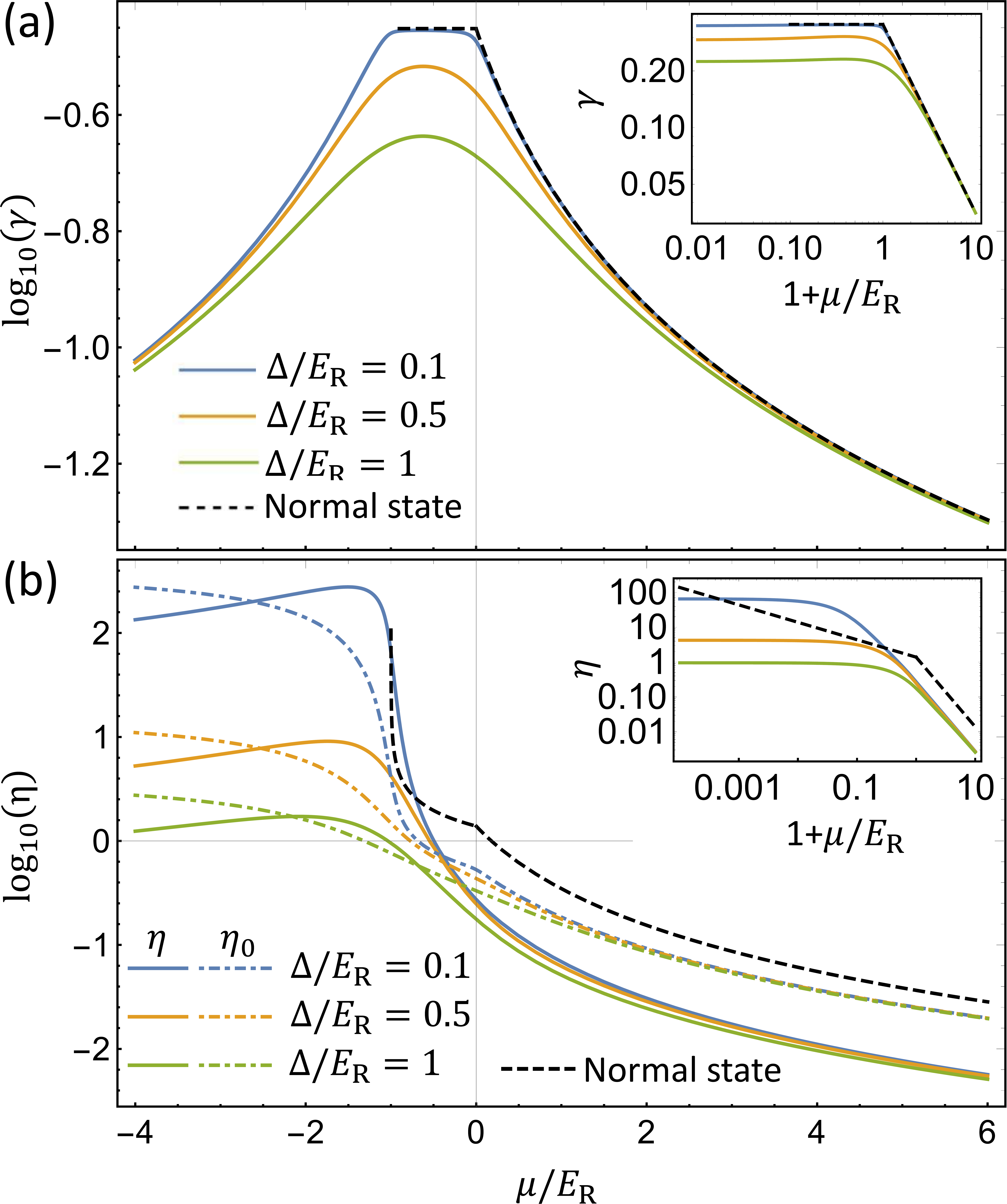}
	\caption{  {\bf Spin density $ M_y $ and spin current density $ j^y $ for a given charge current density $ j_x $.} (a) The logarithm of spin density generation efficiency, defined as, $\gamma=M_y/j_x$  as a function of the chemical potential $\mu$ for various values of the superconductivity order parameter $\Delta$. The black dashed curve is for the normal state.  (b) The logarithm of spin current generation efficiency, defined as  $\eta=j^y/j_x^2$, as a function of the chemical potential $\mu$ for various values of the $\Delta$. The dashed colorful curves are for the quantity $ \eta_0 =  j^y_0 / j_x^2 $ where $ j^y_0 =  (j_x/e) (M_y/n_e) $ denotes a direction product of the particle current $ (j_x/e) $ and the spin polarization per particle $ (M_y/n_e) $. $ n_e $ denotes the electron density and $ e $ is the electron charge. The same color corresponds to the same value of $ \Delta $. The insets are the log-log plot of $ \eta $ emphasizing the regime with negative $ \mu $. }
	\label{FIGratio}
\end{figure}

Since $j_x \sim q_x$, we define two coefficients 
\begin{align}
\gamma=M_y/j_x, \textbf{    }
\eta=j^y/j_x^2,
\end{align}
which denote the efficiency of spin density and spin-current generation, respectively, for given charge current density. They are shown in FIG.\ref{FIGratio}. For large $ \mu $, both the spin density and spin current decrease in power-law, $ \gamma \sim (\mu+E_{\text R})^{-1} $, $ \eta \sim (\mu+E_{\text R})^{-2} $. When $ \mu $ goes below zero, both $ \gamma $ and $ \eta $  increase. Interestingly, if $ \Delta $ is small compared to Rashba splitting, $ \gamma $ stays constant for $ \mu <0 $ until it reaches the band edge, below which it decreases again. For spin current, the ratio $ \eta $ keep increasing before $ \mu $ reaches the band edge. Below that, $ \eta $  decreases very slowly. That means the efficiency of spin current generation in the BEC regime is very high.

\subsection*{ Relation between spin density and spin current}
As shown above, the functional behaviors of the spin density and spin current are similar in some parameter regions but very different in others. To further clarify the mechanism of the spin-current generation, we investigate its relation to the spin polarization. 

When spin polarization is induced by supercurrent, we may intuitively expect a spin current 
\begin{align}
j^y_0 = (M_y/n_e) (j_x/e)= \eta_0 j_x^2, \label{EqRel0}
\end{align} 
which is just the particle current times the spin polarization of each particle. 
Using Eq.(\ref{EqLondon}$-$\ref{EqRel0}) and Eq.(\ref{EqMy}), we get $\eta_0=\gamma e/n_e$ and
$
\gamma = \frac{  m \alpha \hbar }{4 e E_{\text R}}\frac{g(d,u)}{\pi\hbar^2 n_e(u)/(m E_{\text R})-  g(d,u)} .
$
In FIG.\ref{FIGratio}(b), by comparing $ \eta $ (solid curves) with $ \eta_0 $ (dashed curves in corresponding colors), we find  
that the differences between them are not small in general. Thus, Eq.(\ref{EqRel0}) does not fully describe the origin of the spin current because the spin and momentum degrees of freedom cannot be separated in systems with SOI. 

\subsection*{ Comparison to normal states}
In normal dissipative Rashba systems, spin density \cite{EdelsteinNormal} and spin current \cite{Yu,Hamamoto} can also be generated when an external electric field $ \mathbf E $ is applied. In such a case, the coefficients $ \gamma $ and  $ \eta $ are shown by the black dashed curves in FIG.\ref{FIGratio}. It turns out that the spin polarization efficiency for SC state and that for  normal state are almost the same if $ \Delta $ is small. For spin current generation, $ \eta $ for the SC state is smaller than the normal state in most of the parameter regime. Only when $ \mu $ is very close to the band edge and $ \Delta $ is small, the SC state shows a higher efficiency. The comparison of the normal state and the SC state becomes clearer in the log-log plots as shown in the insets of FIG.\ref{FIGratio}.	When $ \mu<-E_{\text R} $, it is not possible to pass a current in the normal state at $ T=0 $. However, there could be BEC SCs in this limit, which can carry supercurrent and gives nonzero spin densities and spin currents.  

Although a dissipative normal system seems actually better at generating spin current than its SC counterpart for most of the parameter space, we should note that, in the normal state,  the spin current, as well as the charge current, is generated by an electric field  and suffers from dissipation and heating, making it impossible to maintain long-range coherence. On the other hand, 	the spin current in the SC state has very low dissipation (if any) because it is purely due to the flow of a supercurrent. In other words, it is the property of the Cooper pair condensate and long-range coherence is guaranteed, making it a spin supercurrent \cite{JacobNP,Linder2019,Eschrig1,Eschrig2}.

It should be emphasized that the spin current in a coherent system does not necessarily correspond to the coherent spin current as defined by Leurs et al. \cite{Zaanen} as mentioned previously.  
In the formalism of non-Abelian gauge fields, the spin current operator can be decomposed into two parts, $ \hat{\mathbf J}^a=\hat{\mathbf J}_{\text NC}^a+\hat{\mathbf J}_{\text C}^a $. The first term $ \hat{\mathbf J}_{\text NC}^a=\hat{\mathbf J} \hat{S}^a $ is a direct product of the charge current operator and the spin operator and is called non-coherent. The second term $ \hat{\mathbf J}_{\text  C}^a $ involves spin entanglement, according to Leurs at al., and is called coherent. In our calculation, the spin current operator is actually $ \hat{\mathbf J}_{\text NC}^a $. 
Such spin supercurrent carried by Cooper pairs does induce spin transport and may be used to create new spintronic devices with long-range coherence. 

\section*{Discussion}
We have shown that a supercurrent in an superconductor with Rashba SOI can induce spin density and spin current. Let's take the interface between a LaAlO$_3$ and a SrTiO$_3$ as an example and estimate the realistic magnitude of these effects. The critical current density (2D) is $ \sim 10^{-3} A/cm $, effective mass  $ m^*=1.5m_e $ and the super-fluid density  $ n_s \sim 10^{13}cm^{-2} $ \cite{Golubev}. Assuming $ E_{\text R} \sim 10meV $ ($ \alpha \sim 10^5 m/s $) and replacing the electron density $ n_e $ by the super-fluid density $ n_s $, the second term in Eq.(\ref{EqLondon}) has the same order of magnitude as the first term when the Fermi level is high, i.e. or $ u \gg 1 $. For a super-current density of $ j_x \sim 10^{-5} A/cm $, which is about $1/100$ of the critical current density, the corresponding $ q_x \sim 10^3 m^{-1} $. Then, according to Eq.(\ref{Eqjp}) and Eq.(\ref{EqMy}),  the spin current density $ j^y \sim  10^8 \hbar \cdot s^{-1}  cm^{-1}$ and spin density $ M_y \sim 10^{7} \hbar \cdot cm^{-2}$. Converted to charge current by replacing $ \hbar/2 $ by $ e $, such a spin current corresponds to an charge current density of $ \sim  0.01 nA/cm $. This is rather small. However, If the Fermi level is decreased, say by gating \cite{Eerkes,Caviglia1,Triscone,Shalom}, and it is still superconducting, the spin current can be enhanced by several orders of magnitude as shown in FIG.\ref{FIGratio}(b). In that case, it should be detectable by inverse spin Hall effect \cite{Tatara}, by connecting the superconductor with a light-emitting diode \cite{Molenkamp}, or by other methods. Our results may be generalized to other spin-orbit coupled systems where general properties of the Edelstein effect have been recently studied \cite{Wenyu2019}.

\begin{figure}
	\includegraphics[width=3.2in]{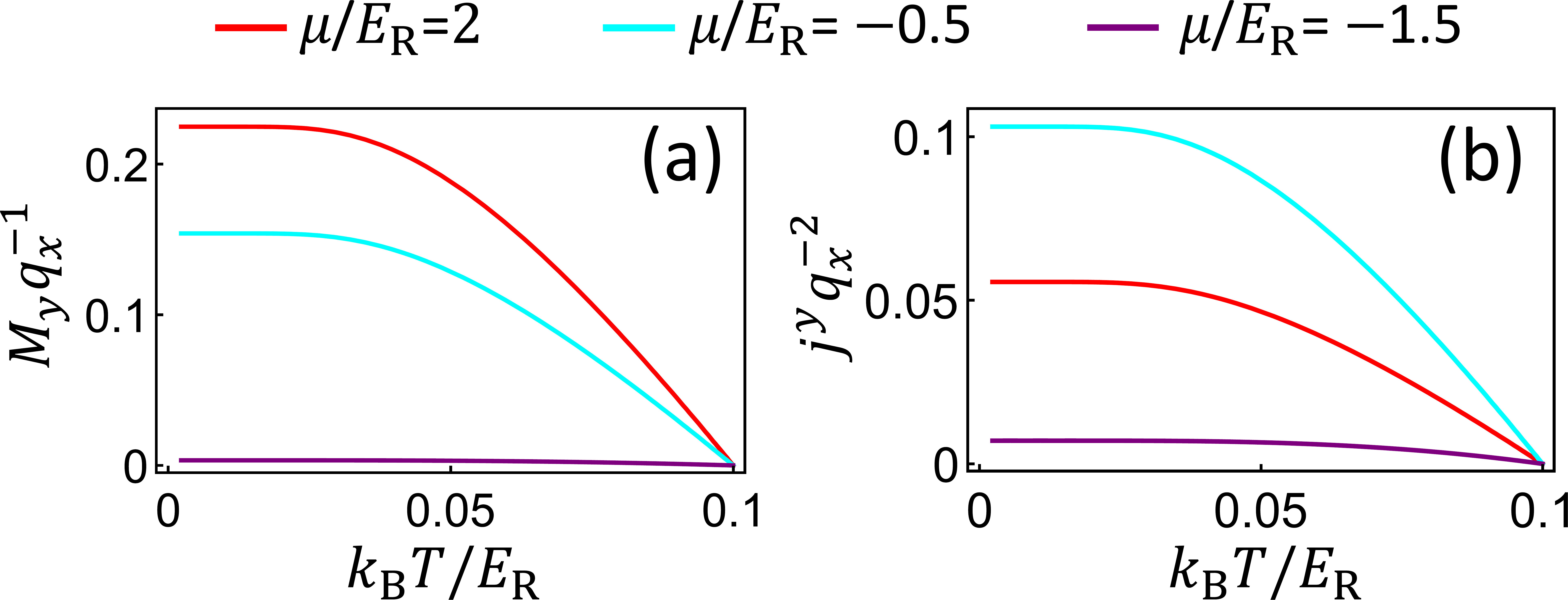} 
	\caption{ {\bf Temperature effect.} (a) The spin density (with a given Cooper pair wave vector $ q_x $) as a function of temperature for various values of the chemical potential $ \mu $. The Rashba splitting energy $ E_{\text R} $ is used as the energy scale. The superconductivity critical temperature is set to be $ T_{\text c}=0.1 E_{\text R}/k_{\text B}  $ with $ k_{\text B} $ being the Boltzmann constant. (b)  The spin current density (with a given $ q_x $) as a function of temperature for various values of the chemical potential $ \mu $.}
	\label{FIG_T}
\end{figure} 

Another important issue is the effect of finite temperature which affects the results through both the Fermi distribution of the quasi-particles and  change of  $ \Delta $. In the following, we assume that the function $ \Delta(T) $ is determined by the BCS theory. FIG.\ref{FIG_T}(a) shows the spin density as a function of temperature for different values of $ \mu $. Generally, the spin density starts to grow linearly when $ T < T_c $, and it saturates when $ T \rightarrow 0 $. It is exactly the same feature as the superfluid density $ \rho_s$ in standard BCS theory. This is no surprise since Eq.(\ref{EqLondon}) indicates that the spin density is part of the supercurrent. When the Fermi level moves downward, the temperature dependence remains the same. 
In FIG. \ref{FIG_T}(b), we show the temperature dependence of the spin current.  
Further discussion about temperature effect can be found in Supplementary Note 3. The temperature dependence in the case of normal state is shown in Supplementary Note 4 for comparison. The spin current generation efficiency as a function of chemical potential is obtained for finite temperature in Supplementary Note 5 which shows that our conclusions are still valid. 

One may also note that this system is a paradigm for the study of topological superconductors. In fact, an out-of-plane Zeeman field can drive it into a topological phase with Majorana edge modes if $ \mu \sim 0 $. Then  the following questions become interesting: (1) What is the spin current contributed by the edge modes, if any? (2) How are topological phase transitions affected by the supercurrent? In fact, the second question has recently been discussed  but in a quasi-one-dimensional system where it is found that a supercurrent shifts the topological phase transitions \cite{Klinovaja}. Similar things may happen for two-dimensional systems as well. In this way, one may control  topological phases by a supercurrent, achieving new methods for coherent quantum-device manipulation.

To conclude, we have studied the spin and spin current generation
by the supercurrent in two-dimensional noncentrosymmetric superconductor
with Rashba spin-orbit interaction. The spin degrees of freedom
are partially activated due to the noncentrosymmetry even when
on-site pairing between up and down spins is considered.
When the chemical potential is below the band crossing point and
approaching to the
Band edge, i.e., in dilute electron density limit, the large enhancement of the
spin supercurrent generation occurs although the spin density is small.
The efficiency of the spin supercurrent does not decrease so much even when the
chemical potential is below the band edge, i.e., in the BEC limit of the
superconductivity.
Furthermore, the carrier density can be controlled there by gating, and
the chemical potential dependence can be studied experimentally.
These studies will pave a route toward the dissipationless
superconducting spintronics,
where the transfer of spin information without the energy loss is possible
through the long range quantum coherence of the system.
	
\section*{Methods}

\subsection*{ Non-Abelian gauge fields and Rashba spin-orbit interaction } 
The Hamiltonian with SU(2) gauge symmetry of free electrons in magnetic field is
\begin{align}
	&\hat{H}_N = \sum_{\mathbf k} c^\dagger_{s_1,\mathbf k} H_{s_1,s_2}(\mathbf k)c_{s_2,\mathbf k},\\
	&H(\mathbf k) = \frac{1}{2m} (\hbar {\bf k} + e {\bf A} -\bar{g}{\bf W}^\nu \sigma^\nu)^2 + \mu_{\text B} B^\nu \sigma^\nu-\mu,
	\end{align}
	where $\mathbf A$ is the U(1) gauge potential of electromagnetic field and $\mathbf W^\nu$ is the SU(2) gauge potential. The constant $ \mu_{\text B}  $ denotes the Bohr magneton and $\sigma^{\nu=1,2,3}$  are Pauli matrices. The coefficient $\bar{g}$ is a coupling constant to be assigned later. 
	Rashba SOI corresponds to the existence of the following SU(2) gauge field \cite{Tokatly}, 
	\begin{align}
	\bar{g}\bar{W}_i^\nu = \frac{e\hbar}{mc^2}\epsilon_{ij\nu} E_j.
	\end{align}
	Note that we have introduced the speed of light $ c $ and the Levi-Civita symbol $ \epsilon_{ij\nu} $.   The Hamiltonian with this gauge field can be rewritten as
	\begin{align}
	H({\bf k})=&\frac{1}{2m} (\hbar {\bf k}+ e {\bf A})^2 +  \frac{1}{2m}  (\bar{W}_j^\nu \sigma^\nu \bar{W}_j^\gamma \sigma^\gamma)\notag\\
	&- \frac{1}{m} [(\hbar k_i+e A_i) \bar{W}_i^\nu \sigma^\nu] + \mu_{\text B} B^\nu \sigma^\nu-\mu,
	\\
	= &\frac{1}{2m} (\hbar {\bf k} + e {\bf A})^2 + \frac{e \hbar E_j}{m^2c^2} (\hbar k_i+e A_i) \epsilon_{j i\nu} 
	\sigma^\nu \notag\\
&	+ \mu_{\text B} B^\nu \sigma^\nu 
	 - \mu+  \frac{1}{2m} \bar{W}^\nu_j \bar{W}^\nu_j,
	\\
	= &\frac{1}{2m} (\hbar {\bf k} + e {\bf A})^2 + \frac{e \hbar}{m^2c^2} {\bf E}\cdot(\hbar {\bf k}+e {\bf A})\times {\bf \sigma} \notag\\
	&+ \mu_{\text B} B^\nu \sigma^\nu - \mu+  \frac{1}{m} (\frac{e\hbar}{mc^2})^2 |{\bf E}|^2.
	\end{align}
    The second term describes the Rashba SOI. So, the Rashba SOI will be included by assuming a constant SU(2) gauge field $\tilde{\bf W}^\nu$ and the SU(2) gauge symmetric Hamiltonian with certain Rashba strength becomes
	\begin{align}
	H_{\text R}({\bf k})= \frac{1}{2m} [\hbar {\bf k} + e {\bf A} - \bar{g}(\delta{\bf W}^\nu+{\bf \bar{W}}^\nu) \sigma^\nu]^2 \\
	+ \mu_{\text B} B^\nu \sigma^\nu-\mu,
	\\
	= \frac{1}{2m} (\hbar {\bf k} + e {\bf A}-\bar{g}{\delta\bf W}^\nu \sigma^\nu)^2 +\alpha (\hbar {\bf k}+e {\bf A})\times {\bf \sigma} 
	\\
	+\mu_{\text B} B^\nu \sigma^\nu - \mu +  \frac{\bar{g}^2}{m}  \delta W^\nu_j \bar{W}^\nu_j +  \frac{1}{m} (\frac{e\hbar}{mc^2})^2 |{\bf E}|^2.
	\end{align}
	The last two terms are just constant energy shift due to the electric field of the Rashba SOI, which we ignore hereafter. Then we have
	\begin{align}
	H_{\text R} ({\bf k})= \frac{1}{2m} (\hbar {\bf k}+ e {\bf A}-\bar{g}{\delta\bf W}^\nu \sigma^\nu)^2\\
	 +\alpha (\hbar {\bf k}+e {\bf A})\times {\bf \sigma}
	+\mu_{\text B} B^\nu \sigma^\nu - \mu .
	\label{hr}
	\end{align}
	In order for the non-Abelian gauge field $ \mathbf W^\nu $ to couple with the spin, we set 
	\begin{align}
	\bar{g}=\hbar/2 .
	\end{align}

	\subsection* { Derivation of  spin density } 
	We calculate the spin density induced by the supercurrent, or by the vector potential $ \mathbf A = (\hbar q_x/e)  \hat{\mathbf x}$  in Eq.(\ref{EqH}), using perturbation method.  The perturbation Hamiltonian $ \hat{H}' $ including both the test fields ($ \mathbf W^\nu $ and $ \mathbf B $)  and the external field ($ \mathbf A $)  is 
	\begin{align}
	& \hat{H}' = \delta\hat{H}_{A1} + \delta\hat{H}_{A2} + \delta\hat{H}_{W} + \delta \hat{H}_{AW} + \delta \hat{H}_B, \label{EqdH}
	\end{align}	
	with 
	\begin{align}
	\delta\hat{H}_{A1}& = \sum_{\mathbf k}[- \frac{\hbar \mathbf k \cdot e \mathbf A}{m}\delta_{ss'} + \alpha e (A_x \sigma^y )_{ss'}]c^\dagger_{s,\mathbf k}  c_{s',\mathbf k},\\
	 \delta\hat{H}_{A2} &= \sum_{\mathbf k}[- \frac{e^2 |\mathbf A|^2}{2 m}\delta_{ss'} ]c^\dagger_{s,\mathbf k}  c_{s',\mathbf k},	\\ 
	 \delta\hat{H}_{W} &=\sum_{\mathbf k} [\frac{-\hbar^2}{m} \mathbf k \cdot  \delta\mathbf W^\nu \sigma^\nu_{ss'} \\
	 & + \alpha \hbar (\delta W_x^y - \delta W_y^x) \delta_{ss'} ]c^\dagger_{s,\mathbf k}  c_{s',\mathbf k},\\
	 \delta \hat{H}_{AW}& = \sum_{\mathbf k}[\frac{e \hbar}{m} \mathbf A \cdot  \delta \mathbf W^\nu \sigma^\nu_{ss'} ]c^\dagger_{s,\mathbf k}  c_{s',\mathbf k}, \\
	 \delta \hat{H}_B & =  \sum_{\mathbf k}  \mu_{\text B} \mathbf B \cdot \mathbf \sigma_{ss'} c^\dagger_{s,\mathbf k}  c_{s',\mathbf k},
	\end{align}	
	which is composed of terms linear and quadratic in $ \mathbf A $ respectively, a term proportional to non-Abelian gauge field $ \mathbf W^\nu $, a term bilinear in $ \mathbf A $ and $ \mathbf W^\nu $, and the term due to the field $ \mathbf B $.  As mentioned previously, the spin density $ M_y $ is linear in $ q_x $ (or $ A_x $) to the lowest order. Consequently,  we should calculate the contribution to the free energy by $ \hat{H}' $ up to the second order. 
	
	At zero temperature, the free energy is simply the ground state energy. 
	\begin{align}
	F_0 = \sum_{\mathbf k,\pm} {E_{h\pm}( k)} = \frac{L^2}{4\pi^2} \int  E_{h\pm}(k)k dk d \theta. 
	\end{align}
	The subscript $ h $ labels the hole bands, i.e. $ E_{h\pm}(k) <0 $. Here we ignored the energy due to Cooper pairs, assuming that the order parameter is not affected by the perturbation term. The unperturbed energy spectrum of the system is
	\begin{align}
	E^0_{e\pm}(\mathbf k)  =\sqrt{\xi_\pm(\mathbf k)^2+\Delta^2},\\ 
	 E^0_{h\pm}(\mathbf k)  = - \sqrt{\xi_\pm(\mathbf k)^2+\Delta^2}. 
	\end{align}
	$	\xi_\pm = \frac{\hbar^2k^2}{2m}-\mu \pm \alpha \hbar k  $ are the energy spectrum of the normal Rashba bands.  
	The lowest order term of the spin density is from the bilinear term $  \sim q_x B_i  $. Then,  the perturbation correction to the free energy is
	\begin{align}
     &\delta F_M = \frac{L^2 \hbar B_y q_x }{16\pi}\times \notag\\ &\int_0^\infty dk  \left(\frac{\xi_-(k)}{\sqrt{\xi_-(k)^2+\Delta^2}}-\frac{\xi_+(k)}{\sqrt{\xi_+(k)^2+\Delta^2}}\right) .
	\end{align}	
	The spin density is 
	\begin{align}
	M_y& = -\frac{1}{L^2}\frac{\partial \delta F_M}{\partial B_y}\notag\\
	& = \frac{ \hbar q_x }{16\pi} \int_0^\infty  dk  \left(\frac{\xi_+(k)}{\sqrt{\xi_+(k)^2+\Delta^2}} - \frac{\xi_-(k)}{\sqrt{\xi_-(k)^2+\Delta^2}}\right).
	\end{align}
	After substituting the expressions of $ \xi_\pm $ into the above integral and defining $ k_\pm = k \pm m \alpha/\hbar  $, it becomes
	\begin{align}
   M_y = &\frac{ \hbar q_x }{16\pi} [\int_{\frac{\alpha  m}{\hbar }}^{\infty +\frac{\alpha  m}{\hbar }} \frac{\frac{k_+^2 \hbar ^2}{2 m}-\left(\mu +E_{\text R}\right)}{\sqrt{\Delta ^2+\left(\frac{k_+^2 \hbar ^2}{2 m}-\left(\mu +E_{\text R}\right)\right)^2}} \, dk_+\notag\\
    &-\int_{-\frac{\alpha  m}{\hbar }}^{\infty -\frac{\alpha  m}{\hbar }} \frac{\frac{k_-^2 \hbar ^2}{2 m}-\left(\mu +E_{\text R}\right)}{\sqrt{\Delta ^2+\left(\frac{k_-^2 \hbar ^2}{2 m}-\left(\mu +E_{\text R}\right)\right)^2}} \, dk_-]\\
   =& \frac{ \hbar q_x }{16\pi} \frac{\alpha  m }{\hbar } [ \int_{1}^{\infty +1} \frac{x^2-u}{\sqrt{d^2+\left(x^2-u\right)^2}}  dx\notag \\
 &  - \int_{-1}^{\infty -1} \frac{x^2-u}{\sqrt{d^2+\left(x^2-u\right)^2}}  dx]
   \\
   = &\frac{ \hbar q_x }{16\pi} \frac{\alpha  m }{\hbar } (2- \int_{-1}^{1} \frac{x^2-u}{\sqrt{d^2+\left(x^2-u\right)^2}}  dx)\\
   =& \frac{m \alpha q_x }{4\pi} g(d,u),
	\end{align}
	with $ u=\mu/E_{\text R}+1 $ and $ d=\Delta/E_{\text R} $.
	
\subsection*{ Derivation of  spin current }
Similar to the spin density derivation, we use perturbation theory to obtain the correction to the free energy $ \delta F_J $. The difference here is that we need to go to the third order perturbation due to the fact that the spin current is quadratic in $ q_x $. After lengthy but straightforward calculation, the free energy correction is
\begin{align}
\delta F_J 
%= \int _0 ^ \infty dk [F_-(\alpha) + F_+(\alpha) ]
=\int _0 ^ \infty dk [F_-(\alpha) - F_-(-\alpha)],
\end{align}
The general form of the function $ F_- $ is complicated. However, in the limit $ \Delta \rightarrow 0 $, it becomes 
\begin{align}
F_-(\alpha) \approx \frac{L^2 q_x^2 W_x^y}{16 \pi^2 }  \{ \frac{5 \pi  \hbar ^2}{8 m} \text{sgn}(\xi -k \alpha  \hbar ) \notag \\ -\frac{\pi k \Delta ^2 \hbar ^3 (4 k \hbar -\alpha  m)}{8 m^2 \left[\Delta ^2+(\xi -\alpha  k \hbar )^2\right]^{3/2}} \}.
\end{align}
By change of integral variable from $ k $ to $ k_- $ to $ x $ similar to the previous calculation, the integral can be written as
\begin{align}
\delta F_J =&  -\frac{L^2 k q_x^2  W_x^y}{16 \pi^2 } \{- \frac{5 \pi  \hbar ^2}{4 m} (k_2 - k_1) \notag\\
&+ \pi  \alpha  \hbar  \int_0^1 \frac{2 d^2 \left( x^2+3/4 \right)}{\left[d^2+\left(x^2-u\right)^2\right]^{3/2}}  dx\notag\\ 
& + \frac{7\pi  \alpha  \hbar }{4} (1+\frac{\mu }{\sqrt{\Delta ^2+\mu ^2}})
\}.
\end{align}
The spin current density is
\begin{align}
j_x^y = -\frac{\partial \delta F_J}{\partial W_x^y} 
=& \frac{ q_x^2 }{16 \pi^2 }  \{- \frac{5 \pi  \hbar ^2}{4 m} (k_2 - k_1) \notag\\
& + \pi  \alpha  \hbar  \int_0^1 \frac{2 d^2 \left(x^2+3/4\right)}{\left[d^2+\left(x^2-u\right)^2\right]^{3/2}}  dx  \notag\\
&+      
\frac{7\pi  \alpha  \hbar }{4} ( 1+\frac{\mu }{\sqrt{\Delta ^2+\mu ^2}})
\}.
\end{align}
We have defined two wave vectors $ k_2 $ and $ k_1 $ which denotes the Fermi wave number of the outer and inner Fermi surfaces respectively. 
\begin{align}
 k_2-k_1= \left\{ 
 \begin{array}{cc}
 2\alpha m /\hbar, &\text{  if } \mu>0,\\
 \frac{2 \sqrt{m \left(2 \mu +\alpha ^2 m\right)}}{\hbar }, &\text{  if }  -E_{\text R}<\mu<0 .
 \end{array} 
 \right. 
\end{align}
When $ \mu $ is large, the integral of the second term is negligible and the first and third terms lead to Eq.(6) of the main text. When $ -E_{\text R} < \mu <0 $, the third term becomes negligible and Eqs.(7)-(8) are obtained.

%\section*{References}	

\bibliographystyle{apsrev4-1}

\section*{Acknowledgment} 
N.N. was supported by Ministry
of Education, Culture, Sports, Science, and Technology
Nos. JP24224009 and JP26103006, the Impulsing Paradigm
Change through Disruptive Technologies Program of Council
for Science, Technology and Innovation (Cabinet Office,
Government of Japan), and Core Research for Evolutionary
Science and Technology (CREST) No. JPMJCR16F1 and No. JPMJCR1874, Japan.

\onecolumngrid

%%%%%%%%%%%  Supplementary  %%%%%%%%%%%%%%%%%%%%

%%%%%%%%%%%%%%%%%%%%%%%%%%%%%%%%%%%%%%%%%%%%%%%%

%%%%%%%%%% Prefix a "S" to all equations, figures, tables and reset the counter %%%%%%%%%%
%\clearpage
\newpage
%\pagebreak

\setcounter{equation}{0}
\setcounter{figure}{0}
%\setcounter{table}{0}
%\setcounter{page}{1}
%\makeatletter

\renewcommand{\theequation}{S\arabic{equation}}
\renewcommand{\figurename}{Supplementary Figure}

%\renewcommand{\theequation}{S\arabic{equation}}
%\renewcommand{\thefigure}{S\arabic{figure}}
%\renewcommand{\bibnumfmt}[1]{[S#1]}
%\renewcommand{\citenumfont}[1]{S#1}
%%%%%%%%%% Prefix a "S" to all equations, figures, tables and reset the counter %%%%%%%%%%

\section*{Supplementary Notes}
%\section*{Supplementary Notes}
\renewcommand{\thesubsection}{Supplementary Note \arabic{subsection}}

\subsection{Triplet Cooper pairs and their polarization}
Since the anomalous Green’s function (or electron-hole propagator) can be written as,
\begin{align}
	G_{eh} (\omega,k)=(\psi_s+\bm d\cdot \bm \sigma)i \sigma_y,
\end{align}
where $ \bm d $ is the triplet component, we have $ \psi_s = \frac{-i}{2} Tr[G_{eh} \sigma_y ], d_x=\frac{-1}{2} Tr[G_{eh} \sigma_z ], d_y=\frac{-i}{2} Tr[G_{eh} ] $  and $ d_z=\frac{1}{2} Tr[G_{eh} \sigma_x ] $. The spin polarization of the Cooper pair is obtained as 
\begin{align}
	S \sim i(\bm d \times  \bm d^* ).
\end{align}
When SOC is present and time-reversal symmetry is preserved, $ \bm S $ vanishes even though $ |\bm d| $ is finite. The SOC induces spinful triplet Cooper pairs. To polarize them, we need to break time-reversal symmetry. In our case, we break the symmetry by driving a supercurrent which is proportional to $ q_x $. In the upper row of Supplementary Figure 1, the Cooper pair polarization is shown in the momentum space for $ q_x\neq 0 $ with  $ \mu \gg E_{\text R} $. (The perpendicular component $ S_z $ vanishes everywhere.) 

\begin{figure}[h]
	\begin{center}
		\includegraphics[width=4.1in]{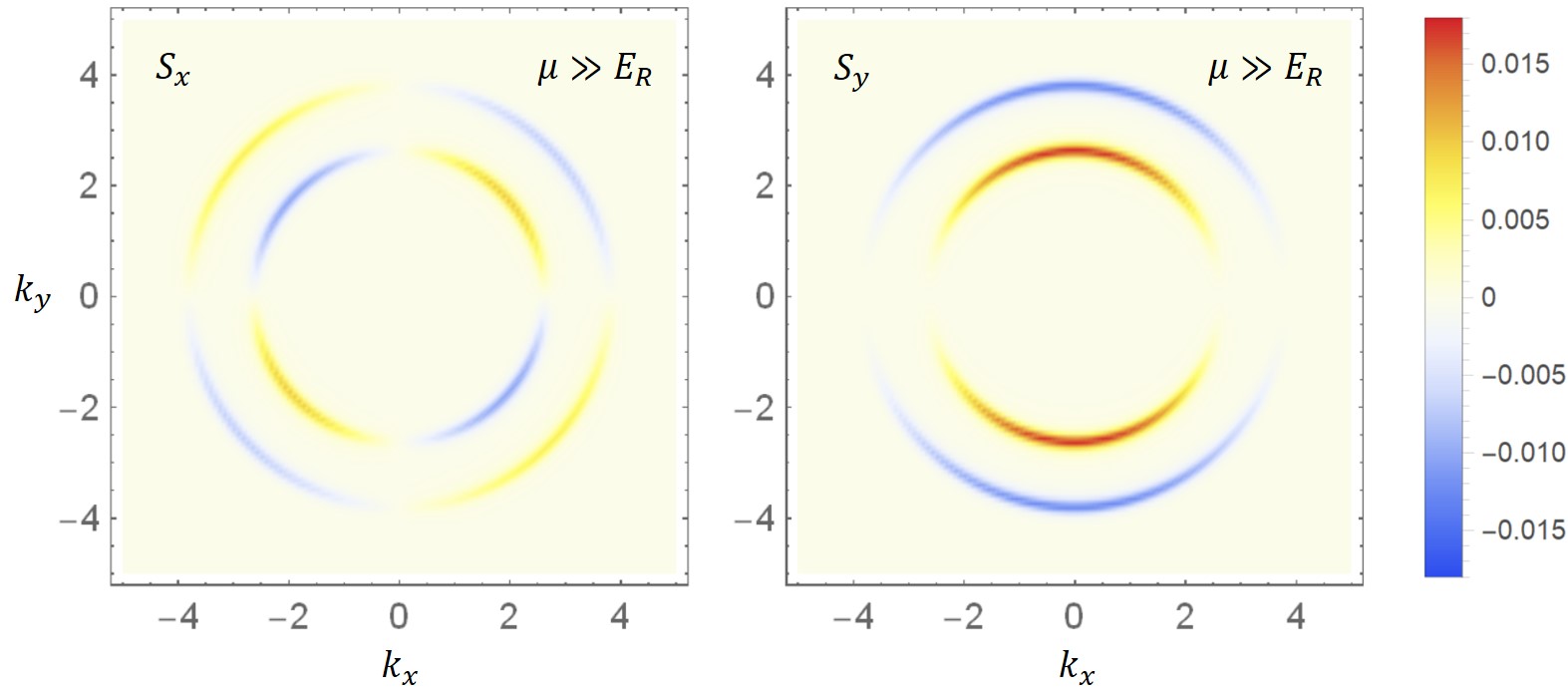}\\
		\includegraphics[width=4in]{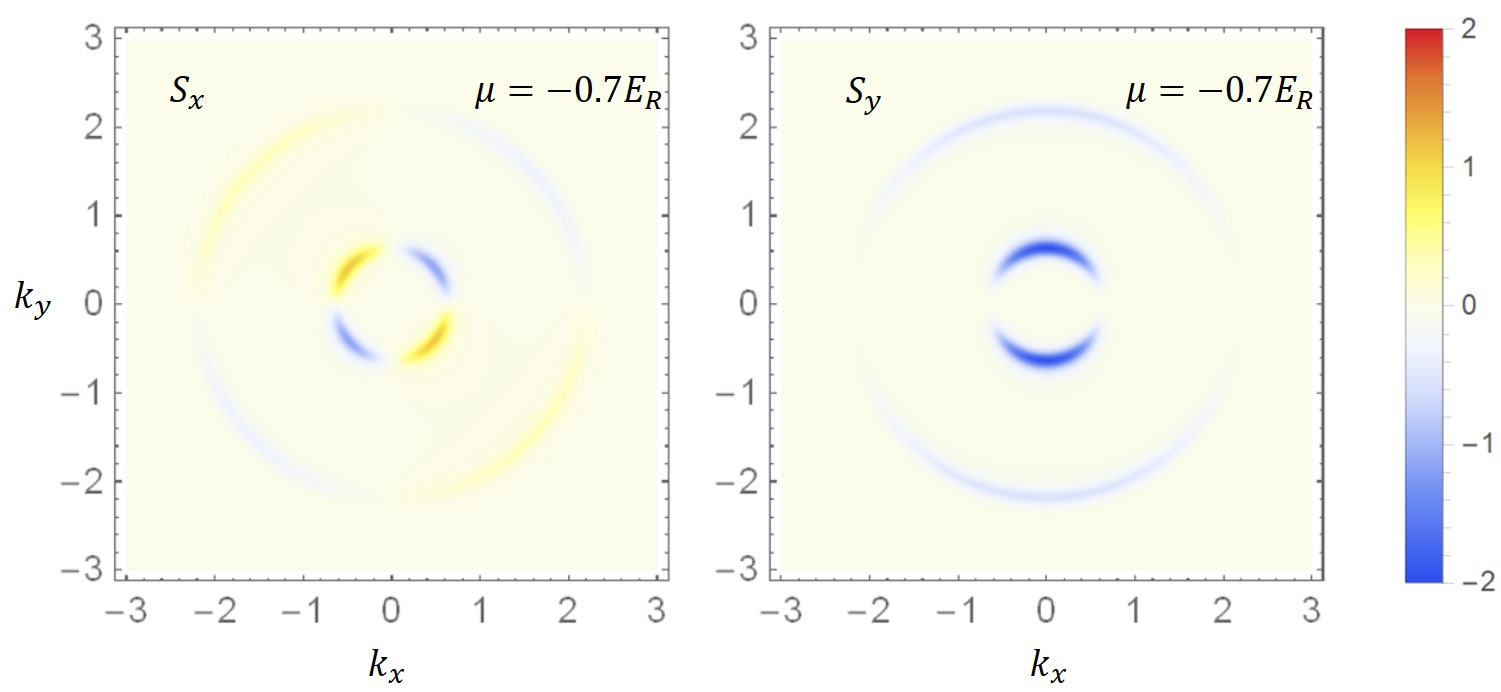}
		\caption{Spin polarization of Cooper pairs in momentum space. $ S_z=0 $ for all $ \bm k $. The perturbation $ q_x $ and other physical constants ($ m, \hbar, e, $ etc.) are set to be unity.}
	\end{center}	
\end{figure}

We see that the two Fermi surfaces (circles) form Cooper pairs with opposite spins. Also, for each Fermi surface, $ S_x $   seems to cancel while $ S_y $ is finite. Actually, $ S_x $ must cancel out due to the mirror symmetry $ M_y $ which changes the signs of $ k_y $ and $ S_x $. As a result, we obtain spin polarization along the y-direction. 

For comparison, we show in the lower row of Supplementary Figure 1 the same quantities with a negative chemical potential $ \mu = - 0.7 E_{\text R} $. Cooper pairs from the inner Fermi circle are now polarized in the same direction as those from the outer circle. 

Since each spin-polarized Cooper pair has a momentum of $ ℏq_x $, they carry a spin current.

\subsection{The discontinuity at $ \Delta = 0 $ when $ T=0 $}
As mentioned in the main text, the coefficients $ j^y/q_x^2 $ and $ M_y/q_x $ are finite when $ \Delta \rightarrow 0 $. On the other hand, they jump to zero at the exact point $ \Delta=0 $ where we thus obtain a discontinuity. However, this does not mean that the quantities $ j^y $ and $ M_y $ are discontinuous at that point because the maximum meaningful value of $ q_x $ actually vanishes  when $ \Delta $ goes to zero. 

\begin{figure}
	\begin{center}
		\includegraphics[width=4in]{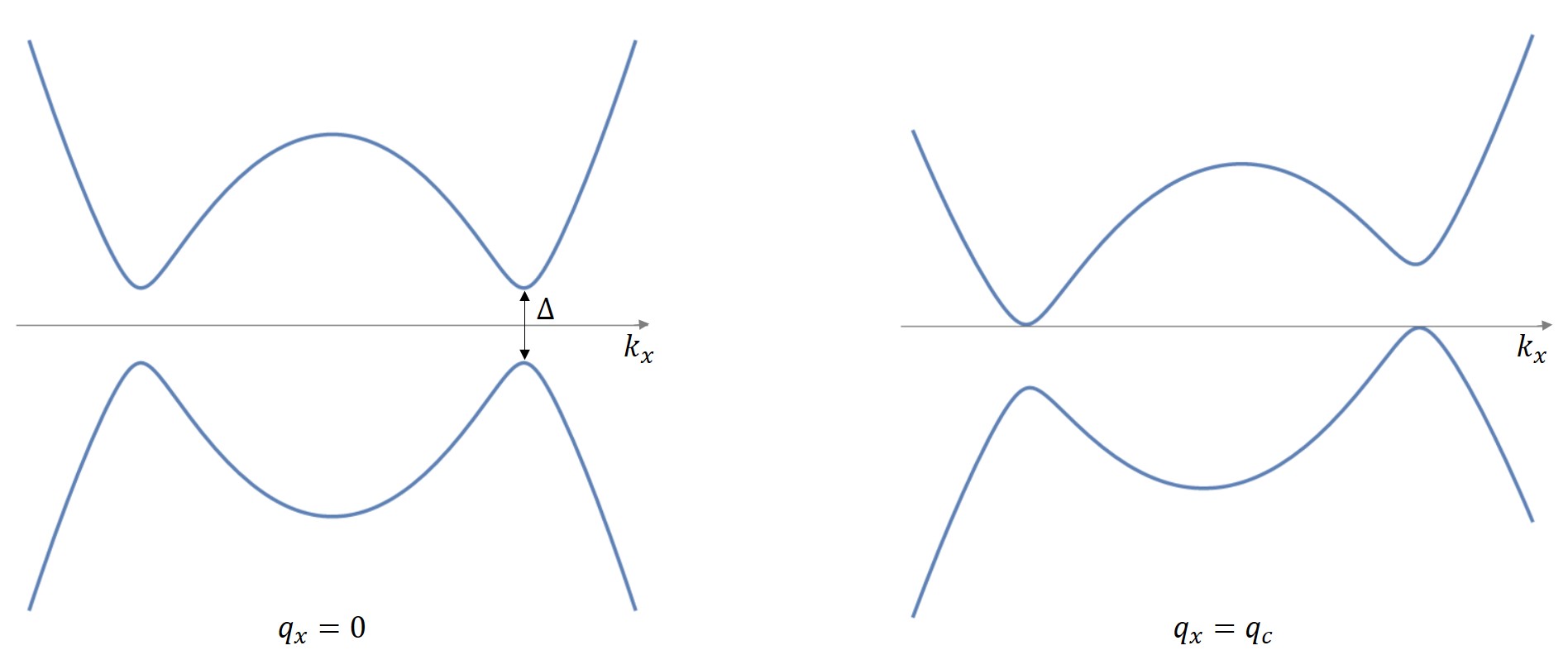}
		\caption{The effect of $ q_x $ (or $ A_x $) on the quasi-particle spectrum. } 	
	\end{center}	
\end{figure}

To understand this, we can look at the effect of $ q_x $, or the supercurrent, on the quasi-particle spectrum. At zero temperature, the perturbation term affects the free energy by changing the energy spectrum of the states with negative energies (assuming unchanged order parameter $ \Delta $). The fact that the quantities $ j^y/q_x^2 $ and $ M_y/q_x $ are almost independent of $ \Delta $ with large $ \mu $ (such as shown in the inset of FIG.1 of the main text) indicates that the free energy is not changed much by varying of the order parameter alone, as long as the spectrum remains gapped. The $ q_x $ tilts the spectrum as shown in Supplementary Figure 2. And when $ q_x \geqslant q_{\text c} $, there is no energy gap and the superconductivity is killed. This $ q_{\text c} $ corresponds to nothing but the critical current of the superconductor. When $ \Delta \rightarrow 0 $, the critical current vanishes  and $ q_{\text c} \rightarrow 0 $. Since $ q_x < q_{\text c} $ must be satisfied, $ q_x $ vanishes when $ \Delta $ approaches zero. Consequently, $ j^y $ and $ M_y $ vanishes as well although $ j^y/q_x^2 $ and $ M_y/q_x $ remain finite. 

\subsection{Further discussion about finite temperature}

Let us first discuss how the discontinuity in the zero-temperature results near $ \Delta=0 $ is affected by finite temperature. As an example, consider the case $ \mu >0 $ in which the spin polarization with finite temperature can be obtained as
\begin{align}
	M_y^+ =\frac{m q_x\alpha}{8\pi} (4-\int dx  \text{ sech}^2[\frac{1}{2} \sqrt{(\frac{\Delta}{k_{\text B} T})^2+x^2}] ).
\end{align}
The spin polarization only depends on the dimensionless parameter $ \Delta/k_{\text R} T $. When $ \Delta=0 $ and $ T\neq 0 $, $ M_y^+(\Delta/k_{\text B} T =0)=0 $, which is consistent with our previous argument for the normal state. When $ \Delta/k_{\text B} T \gg 1 $, it becomes a constant $ M_y^+ (\Delta  /k_{\text B} T \rightarrow \infty) = m \alpha q_x/2\pi $, consistent with Eq.(3). Thus, for $ T\rightarrow 0 $ and small but nonzero $ \Delta $, $ \Delta/k_{\text B} T \rightarrow \infty $ and $ M_y^+ $ is  non-zero, which is the origin of  discontinuity in the zero-temperature results. 

Supplementary Figure 3 show $ M_y/q_x $ and $ J_y/q_x^2 $ as functions of $ \Delta $ for various temperatures and $ \mu =50 $. 
At $ \Delta=0 $, the quantities are actually zero for any nonzero temperature $ T>0 $. And when $ \Delta $ is large, change of temperature has no effect. However, the behavior at small $ \Delta $ is affected. For large temperature, they increase from zero at $ \Delta=0 $ slowly. As temperature decreases, the curve becomes steeper near $ \Delta=0 $. This leads to a infinitely steep curve when $ T=0 $, which is the discontinuity we discussed. This discontinuity is absent when $ \mu<-E_{\text R} $. In this limit, as shown in Supplementary Figure 4, decreasing of temperature does not lead to steeper curves. Instead, they converge to a curve with finite slope. This is true for both spin density and spin current.  

\begin{figure}
	\begin{center}
		\includegraphics[width=5in]{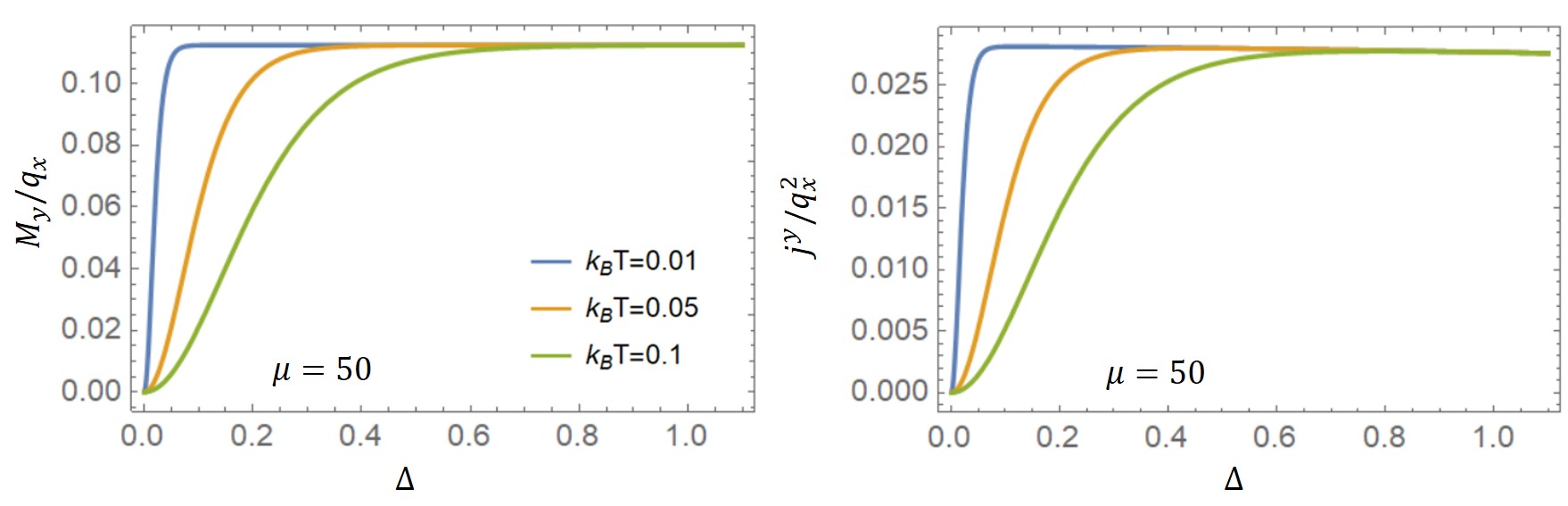}
		\caption{The dependence of $ M_y/q_x $ and $ J_y/q_x^2 $ on the order parameters $ \Delta $ for different temperatures with $ \mu=50 $.}
	\end{center}	
\end{figure}

\begin{figure}
	\begin{center}
		\includegraphics[width=5in]{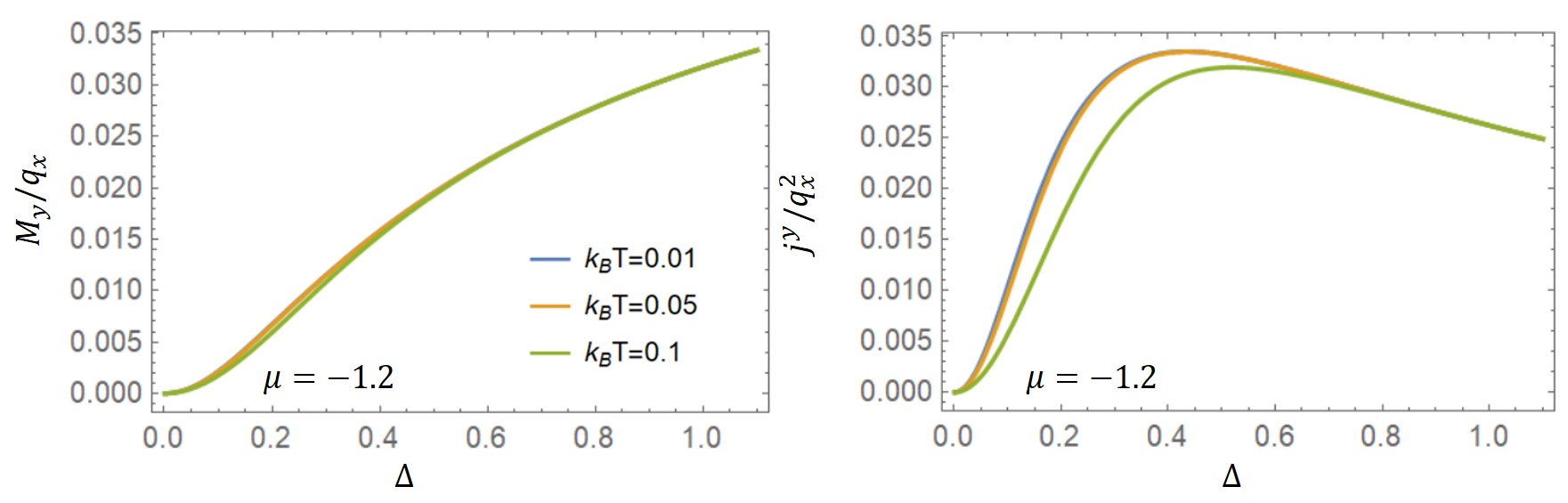}
		\caption{The dependence of $ M_y/q_x $ and $ J_y/q_x^2 $ on the order parameters $ \Delta $ for different temperatures with $ \mu=-1.2 $. }
	\end{center}	
\end{figure}

\begin{figure}
	\begin{center}
		\includegraphics[width=5in]{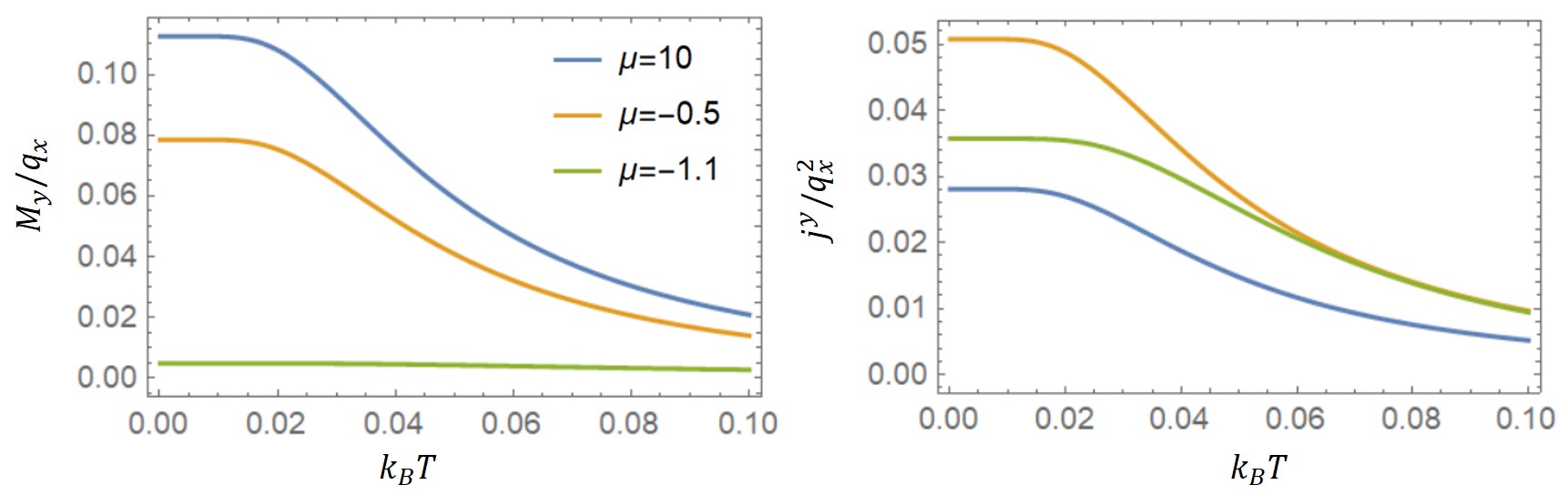}
		\caption{The temperature dependence of $ M_y/q_x $ and $ J_y/q_x^2 $ for fixed $ \Delta=0.1 $ and various values of $ \mu $.}. 
	\end{center}	
\end{figure}

Supplementary Figure 5 shows the temperature dependence of  $ M_y/q_x $ and $ J_y/q_x^2 $ for fixed order parameter. In general, they saturate near $ T=0 $ and decay to zero when temperature increases. Note that the dependence on $ \mu $ of $ j^y $ is nonmonotonic, which can already be seen from FIG. 2(b) of the manuscript. The increase of $ j^y $ as $ \mu $ becomes negative may be due to the change in the density of states, which increases as $ \mu $ approaches the band bottom. When $ \mu $ becomes smaller than $ -E_{\text R} $, the  density of normal states vanishes and $ j^y $ decays.

\subsection{The temperature dependence of the normal state spin density and spin current }

The temperature-dependences of the spin density and spin current of the normal state are shown in the Supplementary Figure 6. For the temperature below the superconductor temperature $ k_{\text R} T_{\text c}=0.1 $, the normal state results are almost independent of T. This is because the temperature in this range is already low enough compared to any other energy scales in the normal state.  Results at higher temperature has been obtained by two of the authors in a previous paper [Hamamoto et al., Phys. Rev. B 95, 224430 (2017)] 

\begin{figure}
	\begin{center}
		\includegraphics[width=4in]{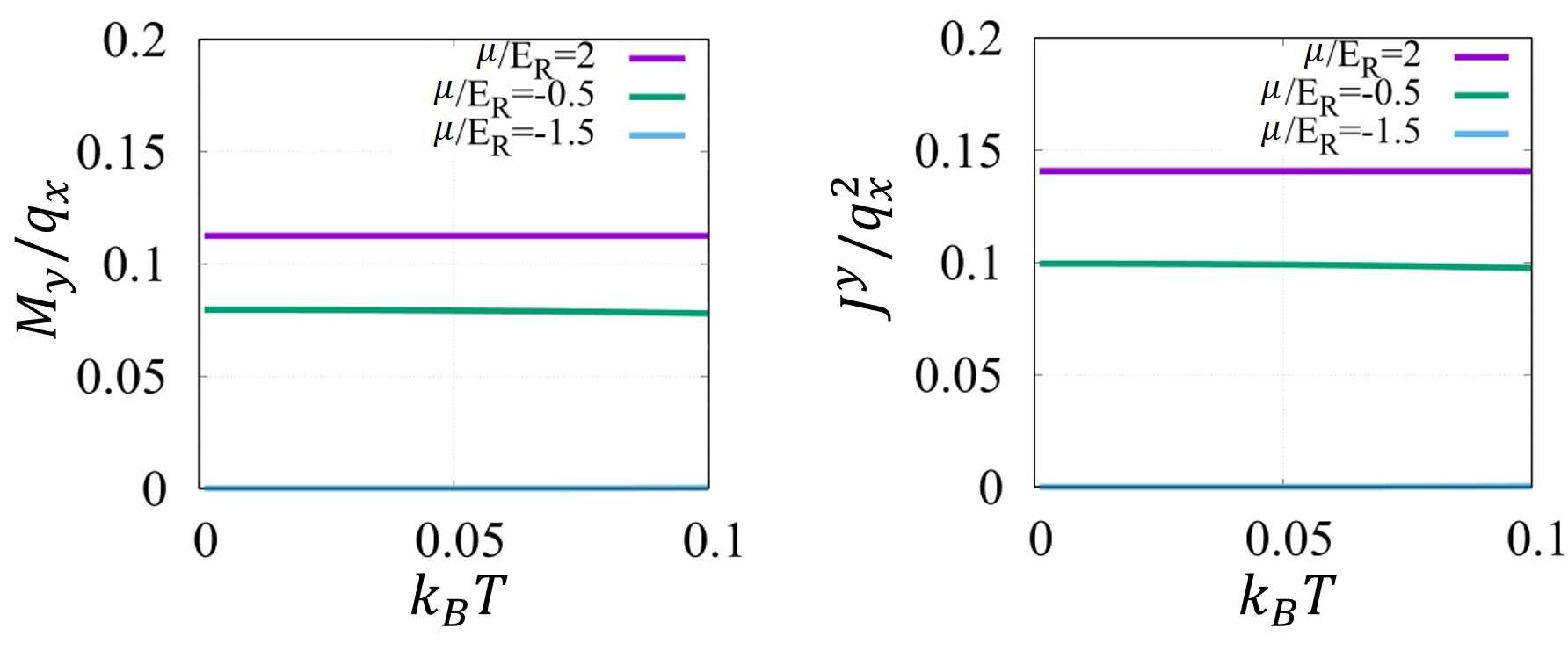}
		\caption{The temperature dependence of the normal state spin density (left) and spin current (right). Note that the wave number $ q_x $ in the normal state is to be understood as the shift of the momentum due to the external electric field.}
	\end{center}	
\end{figure}

\subsection{The comparison of spin current efficiency between superconducting states and normal states at finite temperature}

In Supplementary Figure 7, we compare the spin current efficiency of the superconducting state and that of normal state for various temperatures. For superconducting states, the increase of temperature enhances the spin supercurrent generation efficiency, as long as $ T<T_{\text c} $. This is because the charge supercurrent (the denominator) decreases faster than the spin supercurrent (the numerator) as T increases. The story for the normal state is different, where η is independent of temperature when $ (\mu+E_{\text R})/(k_{\text R} T) \gg 1 $. This is because the temperature scale we consider here is much smaller than any other energy scales of the normal system and it is basically the same as zero-temperature result. When μ is around or below the band bottom, i.e. $ \mu/E_{\text R} \leq -1 $, the temperature dependence becomes clear. Particularly, due to the thermal fluctuation, we obtain finite value even below the band bottom. It is clear from Supplementary Figure 7 that the superconducting spin supercurrent efficiency exceeds that of normal states near the band bottom at non-zero temperatures below $ T_{\text c} $. 

\begin{figure}
	\begin{center}
		\includegraphics[width=3.6in]{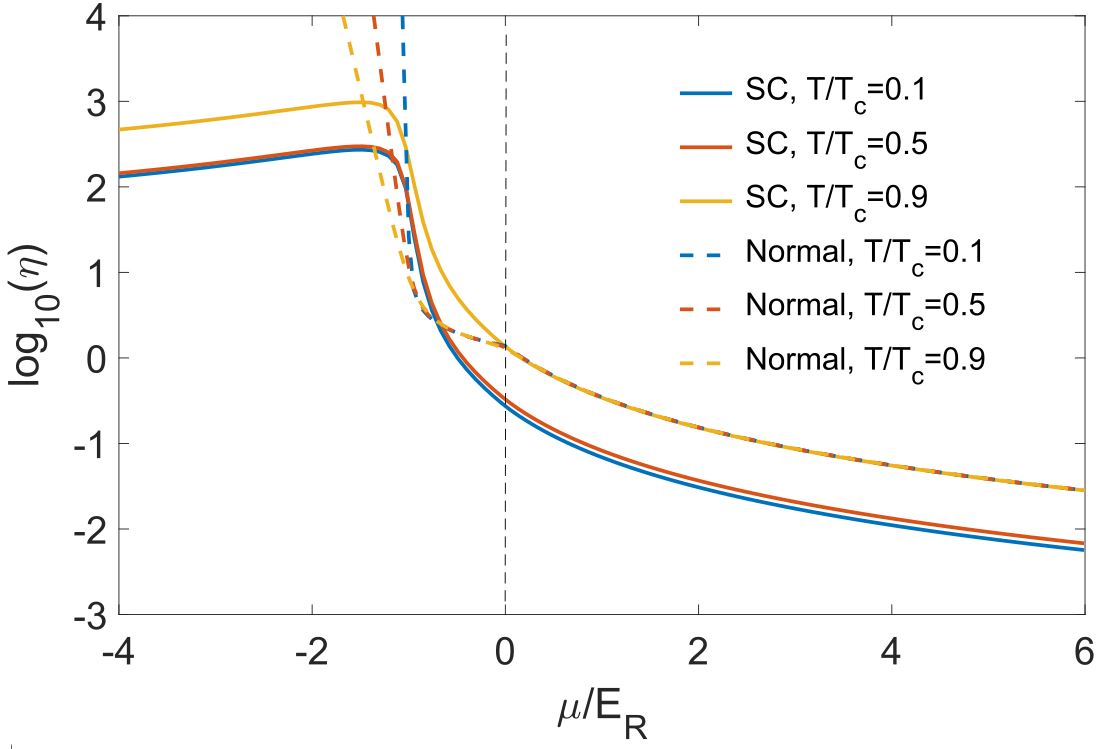}
		\caption{The spin current efficiency $ η=j^y/j_x^2 $ for the superconducting state (solid curves) and for the normal state as functions of chemical potential at various temperatures. We set $ k_{\text R} T_{\text c}=\Delta_{T=0}/1.76=0.057 $ so that $ \Delta_{T=0}=0.1 $ and the blue solid curve here (very small T) approaches the blue solid curve (T=0) in FIG.3(b) of the manuscript. }
	\end{center}	
\end{figure}

\subsection{Dependence on SOC strength}
In Supplementary Figure 8, we show the dependence on SOC strength explicitly ($ \Delta =0.1 $).

The dependence of the spin polarization on the Rashba splitting energy (Left) is monotonic --- larger SOC gives larger polarization. Such monotonic behavior can also be seen in Eq. (5) of the manuscript.   
The dependence of spin supercurrent on SOC strength (Right) is more complicated. We find non-monotonic dependence on SOC strength when $ \mu <0 $. The optimal region is near the band bottom, i.e. $ E_{\text R}\approx |\mu| $. 

\begin{figure}
	\begin{center}
		\includegraphics[width=5in]{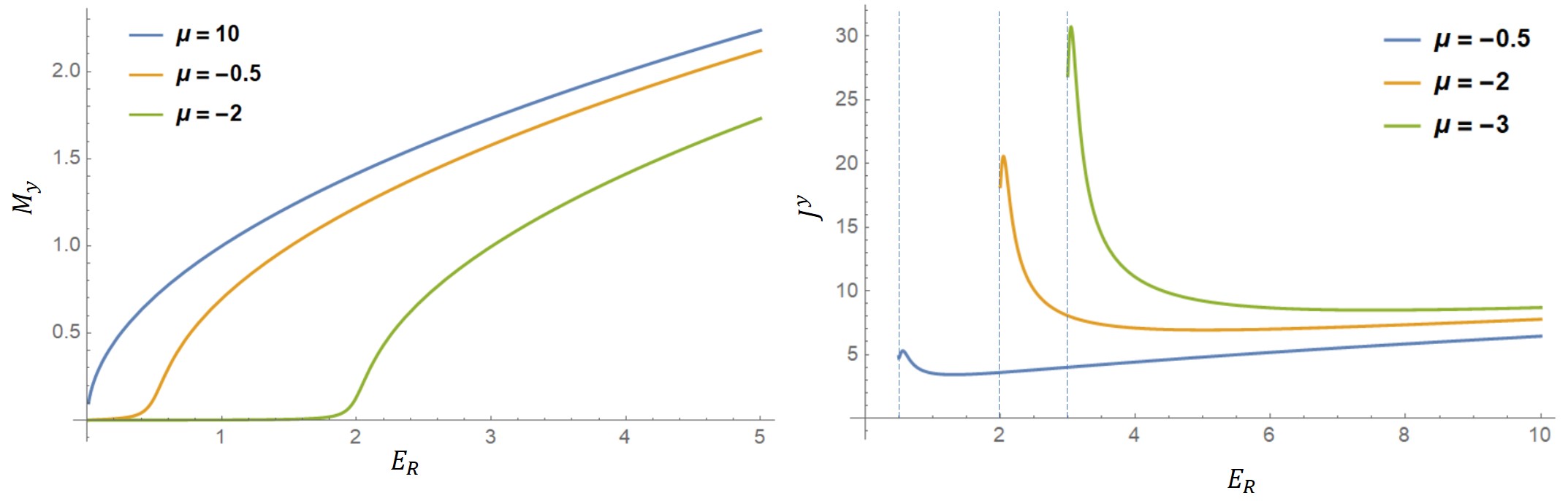}
		\caption{Dependence of $ M_y $ (Left) and $ J^y $ (Right) on SOC strength.}
	\end{center}	
\end{figure}

\end{document}